\documentclass[]{aa} 
\usepackage{graphicx}   
\usepackage{natbib}   
\usepackage{color}   
\bibpunct{(}{)}{;}{a}{}{,} 

\newcommand{\CII}{[\ion{C}{ii}]}   
   
\newcommand{\CI}{[\ion{C}{i}]}   
   
\newcommand{\OI}{[\ion{O}{i}]}

\newcommand{\HII}{\ion{H}{ii}}   
\newcommand{\HI}{\ion{H}{i}}

\newcommand{\msun}{M$_{\odot}$}   
   
\begin{document}   

\title{A complete $^{12}$CO 2--1 map of M51 with HERA: \\ I. Radial 
  averages of CO, \HI, and radio continuum } 
     
   \author{   
     K.F.\,Schuster\inst{1} \and   
     C.\,Kramer\inst{2} \and   
     M.\,Hitschfeld\inst{2} \and   
     S.\,Garcia-Burillo\inst{3} \and   
     B.\,Mookerjea\inst{2,4}    
          }   
   
   \institute{   
     IRAM, 300 Rue de la Piscine, F-38406 S$^t$ Martin d'H\`{e}res, France   
     \and   
     KOSMA, I. Physikalisches Institut, Universit\"at zu K\"oln,   
     Z\"ulpicher Stra\ss{}e 77, 50937 K\"oln, Germany    
     \and   
              Centro Astronomico de Yebes,   
              IGN, E-19080 Guadalajara, Spain         
     \and   
     Department of Astronomy, University of Maryland, College Park, MD 20742, USA  
   }   
   
   \offprints{K.\,Schuster, \email{schuster@iram.es}}   
   \date{Received / Accepted }   
      
   \abstract
   { The mechanisms governing the
       star formation rate in spiral galaxies are not yet clear. The
       nearby, almost face-on, and interacting galaxy M51 offers an
       excellent opportunity to study at high spatial resolutions the
       local star formation laws.}
   {In this first paper, we investigate the correlation of H$_2$, \HI,
     and total gas surface densities with the star forming activity,
     derived from the radio continuum (RC), along radial averages out
     to radii of 12\,kpc.}
   {We have created a complete map of M51 in $^{12}$CO 2--1 at a
     resolution of 450\,kpc using HERA at the IRAM-30m telescope.
     These data are combined with maps of \HI\ and the radio-continuum
     at 20\,cm wavelength. The latter is used to estimate the star
     formation rate (SFR), thus allowing to study the star formation
     efficiency and the local Schmidt law $\Sigma_{\rm
       SFR}\propto\Sigma_{\rm gas}^n$. The velocity dispersion from CO
     is used to study the critical surface density and the
     gravitational stability of the disk.}
   { { The total mass of molecular material derived from
       the integrated $^{12}$CO 2--1 intensities is $2\,10^9$\,\msun.
       The $3\sigma$ detection limit corresponds to a mass of $1.7\,10^5$\,\msun.
       The global star formation rate is 2.56\,\msun\,yr$^{-1}$ and the 
       global gas depletion time is $0.8$\,Gyr.
       \HI\ and RC emission are found to peak on the concave,
       downstream side of the outer south-western CO arm, outside
       the corotation radius.
       The total gas surface density $\Sigma_{\rm gas}$ drops by a
       factor of $\sim20$ from 70\,\msun\,pc$^{-2}$ at the center to
       3\,\msun\,pc$^{-2}$ in the outskirts at radii of 12\,kpc.  The
       fraction of atomic gas gradually increases with radius. The
       ratio of \HI\ over H$_2$ surface densities, $\Sigma_{\rm
         HI}/\Sigma_{\rm H2}$, increases from $\sim0.1$ near the
       center to $\sim20$ in the outskirts without following a simple
       power-law. $\Sigma_{\rm HI}$ starts to exceed $\Sigma_{\rm H2}$
       at a radius of $\sim4$\,kpc.
       The star formation rate per unit area drops from
       $\sim400$\,\msun\,pc$^{-2}$\,Gyr$^{-1}$ in the starburst center
       to $\sim2$\,\msun\,pc$^{-2}$\,Gyr$^{-1}$ in the outskirts.  The
       gas depletion time varies between 0.1\,Gyr in the center and
       1\,Gyr in the outskirts, and is shorter than in other
       non-interacting normal galaxies.
%
       Neither the \HI\ surface densities nor the H$_2$ surface
       densities show a simple power-law dependence on the star
       formation rate per unit area. In contrast, $\Sigma_{\rm gas}$
       and $\Sigma_{\rm SFR}$ are well characterized by a local
       Schmidt law with a power-law index of $n=1.4\pm0.6$.  The index
       equals the global Schmidt law derived from disk-averaged values
       of $\Sigma_{\rm gas}$ and $\Sigma_{\rm SFR}$ of large samples
       of normal and starburst galaxies.
       The critical gas velocity dispersions needed to stabilize the
       gas against gravitational collapse in the differentially
       rotating disk of M51 using the Toomre criterion, vary with
       radius between 1.7 and 6.8\,kms$^{-1}$. Observed radially
       averaged dispersions derived from the CO data vary between
       28\,kms$^{-1}$ in the center and $\sim$8\,kms$^{-1}$ at radii
       of 7 to 9\,kpc. They exceed the critical dispersions by factors
       $Q_{\rm gas}$ of 1 to 5. We speculate that the gravitational
       potential of stars leads to a critically stable disk.
     }}
    {}

 
   \authorrunning{Schuster et  al.} 
   \titlerunning{A complete $^{12}$CO 2--1 map of M51 with HERA}   
   \maketitle   
   
\section{Introduction} 
  
%
%
%
%
%
%
%
  
\begin{figure}[h]   
  \centering   
  \includegraphics[angle=-90,width=8cm]{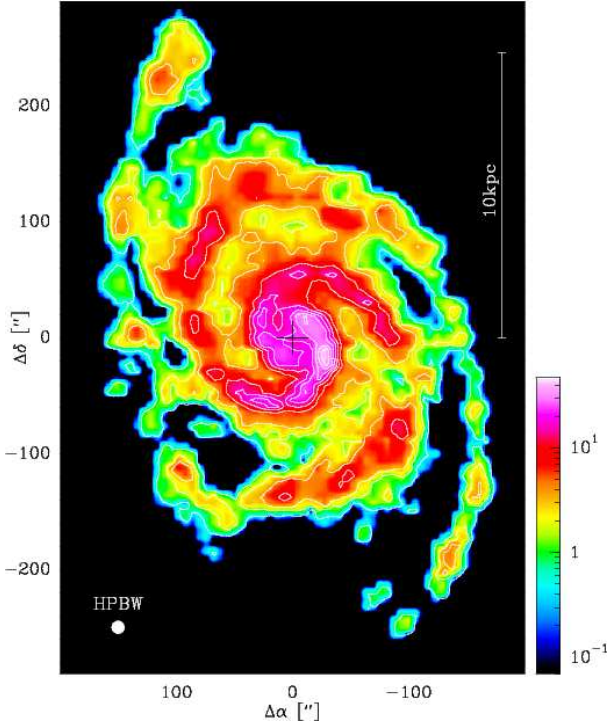}
\caption{Map of $^{12}$CO 2--1 integrated intensities [Kkms$^{-1}$]
showing M51, i.e. NGC\,5194 and its companion galaxy NGC\,5195 in the
north-east.  The image has a resolution of $11''$ and is constructed
from a masked moment calculation \citep{adler1992} to minimize the
noise contribution from emision-free channels when integrating over
the full kinematic extent of M51, 350\,kms$^{-1}$ $<$ v$_{\rm lsr}$ $<
600$\,kms$^{-1}$ .  Color coded intensities are on a logarithmic
scale. The cross marks the 0/0 center position at
$\alpha=13\fh29\fm52\fs7$, $\delta=47\degr11\arcmin43\arcsec$
(eq.2000).
%
The noise level is $1\sigma=0.65$\,Kkms$^{-1}$.   
Contours range in units of 1$\sigma$ from 1, 3, 6, 16 to 56 in steps of 10.  
 The peak intensity is  
47.3\,Kkms$^{-1}$. All intensities are on the $T_{\rm A}^*$ scale.  
%
%
}  
\label{fig_mapco21}   
\end{figure}   
   
 
%
   
  
\begin{figure}[h]   
  \centering 
  \includegraphics[angle=-90,width=8cm]{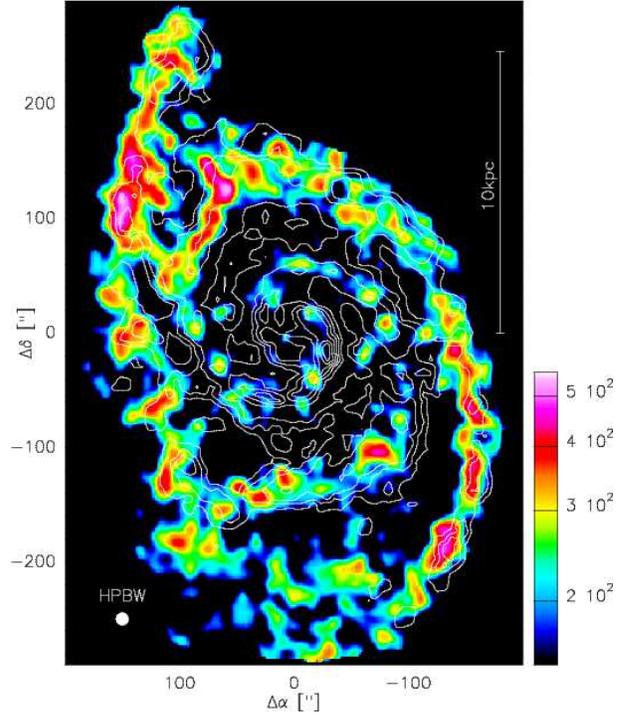}
\caption{VLA map of integrated \HI\ intensities [Jy/beam] at $13''$ resolution 
  \citep{rots1990} in colors.  Contours show integrated $^{12}$CO 2--1
  intensities (cf.\,Fig.\,\ref{fig_mapco21}).
\label{fig_maphi}   
}
\end{figure}   
   
\begin{figure}[h]   
  \centering   
  \includegraphics[angle=-90,width=8cm]{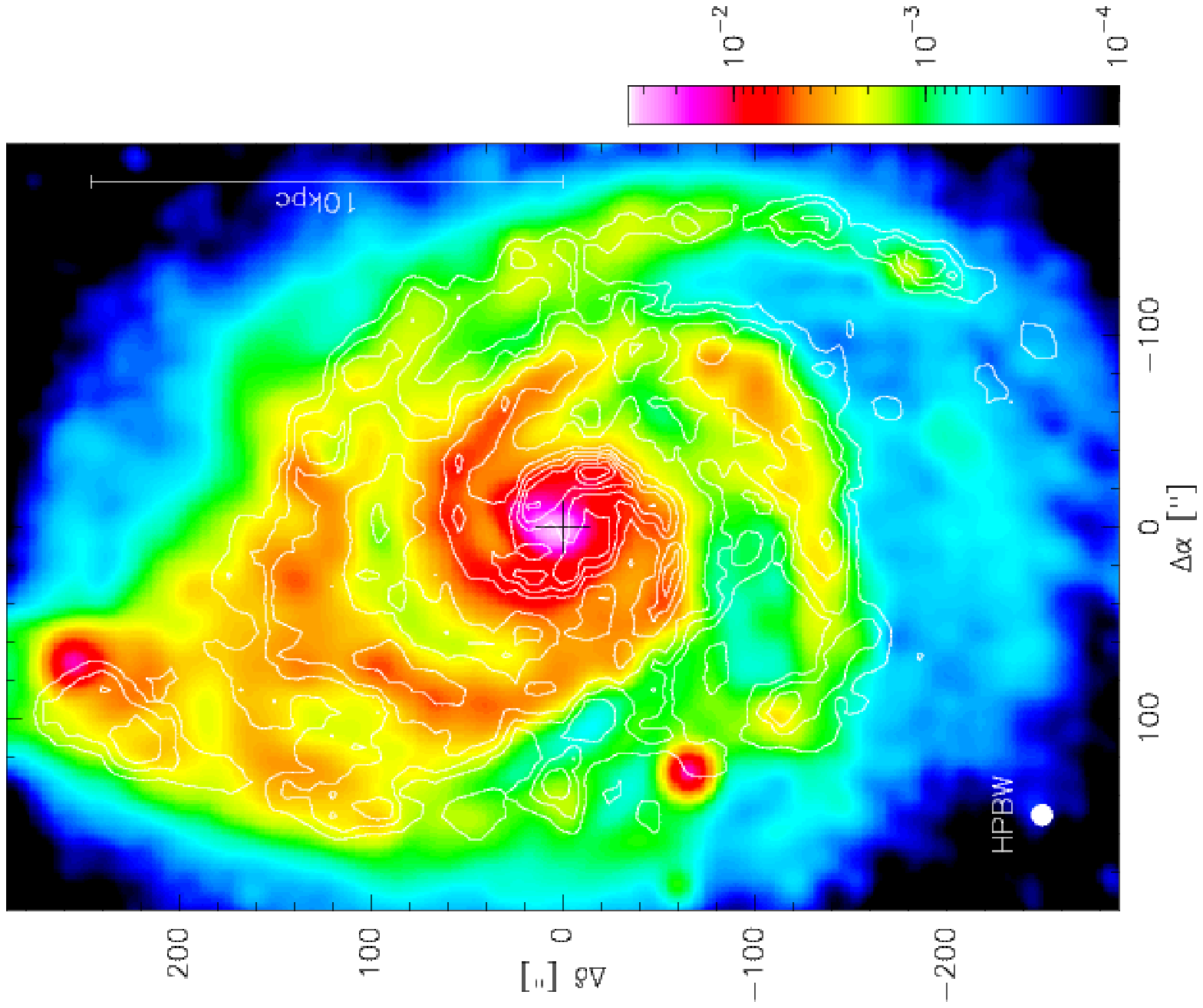}
\caption{Map of radio continuum intensities at 20\,cm 
  (VLA C$+$D, $15''$ resolution) in units of Jy/beam \citep{patrikeev2006}.
  Contours show integrated $^{12}$CO 2--1 intensities
  (cf.\,Fig.\,\ref{fig_mapco21}).
%
\label{fig_map20cm}   
}   
\end{figure}   
   
   
M51 is an interacting, grand-design spiral galaxy at a distance of
only 8.4\,Mpc seen nearly face-on (Table\,\ref{tab_m51_properties}).
It is rich in molecular gas, most of which is found in the two very
prominent spiral arms which are presumably caused by the tidal
interaction 
{
\citep{tully1974c,howard1990}. 
}
Kinematic studies reveal unusually large streaming motions implying a
strong density wave and the presence of galactic shocks
\citep{aalto1999}. The \CII\ map of \citet{nikola2001} shows two
secondary lobes of emission to the northeast and southwest near the
corotation radius of the density wave pattern, presumably due to
cloud-cloud collisions, stimulating star formation.
\citet{calzetti2005} combine
{ Spitzer infrared data of the Infrared Array Camera (IRAC) and the
  Multi Imaging Photometer for Spitzer (MIPS) with UV data from the
  Galaxy Evolution Explorer (GALEX). These data are used}
to discuss the various tracers of star formation in M51. Enhanced star
formation activity at the two lobes seen in \CII\ is probably
triggered by the interaction \citep{nikola2001}.  This is also
indicated by disk simulations of \citet{toomre_toomre1972} which show
tidal tails emerging from the disk at the positions of enhanced star
formation.  VLA observations of the \HI\ line show a prominent tail
outside the main disk extending to the south and east, and covering
more than $25'$, i.e. 100\,kpc, in projected distance \citep{rots1990}
\citep[cf.][]{tilanus_allen1991}.  \citet{meijerink2005} find that a
major fraction of cold dust emission at 850$\,\mu$m stems from an
extended exponential disk with a scale height of 5.45\,kpc, possibly
also tracing total gas column densities. The kinematics of H$\alpha$
emission of M51 were recently studied by the SINGS team
\citep{daigle2006}, improving on the work of \citet{tilanus1991}.

Its proximity and inclination make M51 an ideal target to study the
efficiency at which it forms stars from the gas under the influence
of the interaction, and to compare with the star formation laws
commonly found in spiral galaxies. Spiral galaxies often follow two
empirical laws: 
{ as star formation is fueled by the interstellar gas
  reservoir, the star formation rate (SFR) is proportional to a power
  of the total gas surface density: $\Sigma_{\rm
    SFR}\propto\Sigma_{\rm gas}^n$ .  This appears to hold globally
  when $\Sigma_{\rm gas}$ and $\Sigma_{\rm SFR}$ are averaged over the
  entire star-forming region of a galaxy \citep[][ and
  others]{schmidt1959,kennicutt1998} with $n=1.4$.  This may be
  understood assuming that $\Sigma_{\rm SFR}$ is proportional to the
  gas surface density over the free-fall time $\tau_{\rm ff}$ and that
  the free-fall time and the gas surface density are simple functions
  of the average gas density $\rho$: $\tau_{\rm
    ff}\propto{\rho}^{-1/2}$ and $\Sigma_{\rm gas}\propto\rho$. This
  simple model leads directly to the Schmidt law with an index of 1.5
  \citep[e.g.][]{elmegreen1994}.  More recent studies show that the
  global Schmidt law also appears to hold locally in individual spiral
  galaxies when studying radially averaged profiles of $\Sigma_{\rm
    gas}$ and $\Sigma_{\rm SFR}$ \citep{wong_blitz2002,boissier2003}.
  The second law is that the total gas surface density is of the order
  of a critical gas surface density given by the \citet{toomre1964}
  criterion for gravitational stability
  \citep{martin_kennicutt2001,li2006}.
}

The relative importance of the different components of the
interstellar medium was addressed by \citet{nikola2001} when trying to
explain the mapped \CII\ emission. In a recent study comprising all
major cooling lines of the molecular gas, i.e. those of \CII, \OI, CO, and \CI,
at selected spiral arm positions and the center of M51,
\citet{kramer2005} conclude that the bulk of the emission stems from
clumpy photon dominated regions. Only about 15--30\% of the \CII\ 
emission stems from dense \HII\ regions.  The \CI\ and CO emission of
the center region have also been studied by
\citet{israel2006,israel_baas2002,gerin_phillips2000}.
\citet{schinnerer2004} mapped two distinct regions in the spiral arms
of M51 in HCN, HCO$^+$ and other tracers of the chemistry.

{ \citet{rydbeck2004} presented a large CO 1--0 map of M51 obtained
  at the Onsala-20m telescope.  }
The inner region of M51 were previously mapped in CO 1--0 with the
FCRAO-14m, IRAM-30m, and NRO-45m telescopes by
\citet{scoville_young1983,lord_young1990,gb1993no1,nakai1994,kuno1995,kuno1997}.
These single-dish observations reveal the large-scale emission for
which interferometric observations are not sensitive. OVRO was used to
create CO 1--0 maps by \citet{rand_kulkarni1990} and
\citet{aalto1999}.  \citet{helfer2003} used BIMA in combination with
NRAO-12m single-dish data to map the inner part of M51 in CO 1--0.


Here, we present a complete IRAM-30m CO 2--1 map of M51, improving on
the previous data of \citet{gb1993no1,gb1993no2}.
  
\begin{center}   
\begin{table}[h*]   
\caption[]{\label{tab_m51_properties}   
{\small Basic properties of M51.    
References:    
$^a$ RC3 catalogue \citet{devaucouleurs1991}, 
$^b$ \citet{feldmeier1997}, 
  note that \citet{takats2006} recently reported a distance of only $7.1\pm1.2$\,Mpc,
$^c$ \citet{tully1974b}.   
}}   
\begin{tabular}{lrrrrr}   
\hline \hline   
                             & M51 \\    
\noalign{\smallskip} \hline \noalign{\smallskip}   
RA(2000)                     & 13:29:52.7 \\   
DEC(2000)                    & 47:11:43   \\   
Type                         & SA(s)bc pec$^{a}$ \\      
Distance [Mpc]               & $8.4^{b}$ \\             
$11''$ correspond to         & 448\,pc \\                  
Heliocentric velocity [kms$^{-1}$]  & 463$^{a}$ \\       
Position Angle [deg]         & $170^{c}$ \\                
Inclination [deg]            & $20^{c}$ \\               
%
\noalign{\smallskip} \hline \noalign{\smallskip}   
\end{tabular}   
\end{table}   
\end{center}   

\section{Observations}   
   
Observations of the $^{12}$CO 2--1 emission from M51 were conducted   
with the IRAM-30m telescope in February 2005 using the 18\,element   
focal plane heterodyne receiver array HERA \citep{schuster2004}   
together with the WILMA autocorrelator backend.  WILMA has a channel   
spacing of 2.6\,kms$^{-1}$ (2\,MHz) and a bandwidth of   
1200\,kms$^{-1}$ (930\,MHz). Pixel 2 of the second HERA polarization   
showed excess noise and was ignored from further analysis.   
   
Observations were conducted in position switched on-the-fly (OTF) mode
scanning M51 in right ascension. Sampling was $6''$ in RA. HERA was
rotated by $18.5\degr$ to obtain a spacing of $7.6''$ in declination
between adjacent scan lines \citep{schuster2004}. This corresponds to
near Nyquist sampling for a half power beamwidth (HPBW) of $11''$.
The resulting map has a size of $11'\times11'$.
Figure\,\ref{fig_mapco21} shows the $7'\times10'$
sub-region where CO was detected.  An emission-free reference position
was selected at offsets ($10'$,$0'$).  The mean baseline rms is 18\,mK
at 5\,kms$^{-1}$ velocity resolution on the $T_{\rm A}^*$ scale. To
correct to first order for the telescope efficiencies, i.e. to go from
the $T_{\rm A}^*$ scale to the $T_{\rm mb}$ scale, we simply
multiplied the antenna temperature data with the ratio of forward
efficiency $F_{\rm eff}=0.91$ over main beam efficiency $B_{\rm
  eff}=0.52$.  These numbers show that the mapped spatial structure of
M51 is to some extent smeared out by the error beams of the IRAM-30m
telescope \citep{greve1998}. All data reduction was done using the
GILDAS\footnote{\tt http://www.iram.fr/IRAMFR/GILDAS} software package
supported at IRAM.

%
   
\section{Data} 
\subsection{Molecular gas distribution} 
\label{sec-molecular-gas-distribution}   

  
The HERA map of $^{12}$CO 2--1 (Fig.\,\ref{fig_mapco21}) is the first
CO map of M51 encompassing the companion galaxy as well as the
south-western arm out to a radius of $\sim12\,$kpc in a homogeneously
sampled data set at linear scales of down to 450\,pc.
   
The emission detected with the 30m telescope traces the well known
two-armed spiral pattern out to the companion galaxy NGC\,5195, which
shows up brightly in the north-east at $\sim10.5$\,kpc radial
distance, and out to the south-western tip of the second arm at the
opposite side of M51.  The outer parts of the two arms in the west and
in the east appear more fragmented than the inner parts.  The western
arm especially is almost unresolved.  Inter-arm emission is detected
above the $3\sigma$ level out to radii of about 6\,kpc.  Several
spoke-like structures connect the spiral arms radially.
 
{ To estimate the total H$_2$ column densities from the
  integrated CO 2--1 intensities, we assume a 2--1/1--0 intensity
  ratio of 0.8 as found by \citet{gb1993no1}. \citet{gb1993no1} and
  \citet{guelin1995} independently derived the CO-to-H$_2$ conversion
  factor $X$, for M51 from CO 1--0 and dust continuum data. They find
  that it is a factor 4 smaller than the Milky Way value $X_{\rm MW}$
  \citep{strong1988,strong_mattox1996} and constant with radius.
  Similar X-factors for M51 were found by \citet{nakai_kuno1995},
  using CO data and visual extinctions towards \HII\ regions, and by
  \citet{adler1992} using BIMA CO 1--0 data assuming that the GMCs are
  in virial equilibrium.

  
  For the total H$_2$ column density per beam in M51, we use: $N({\rm
    H}_2)=0.25\,X_{\rm MW}\,(1/0.8)\int T_{\rm mb}({\rm CO}(2-1)) dv$
  with $X_{\rm MW}=2.3\,10^{20}$cm$^{-2}$(Kkms$^{-1})^{-1}$.  }


{ The X-factor may be a function of the metallicity in spiral
  galaxies as has been suggested by several authors
  \citep[e.g.][]{arimoto1996}. However, other factors like the
  radiation field or the cosmic-ray rate also have a strong impact on
  the CO-to-H$_2$ conversion factor \citep{bell2006}. The
  metallicities of M51 have been found to be slightly supersolar
  showing only a shallow drop with radius by only
  $-0.02$\,dex\,kpc$^{-1}$ \citep{bresolin2004}: $ 12+\log({\rm O/H})
  = 8.72 (\pm 0.09) - 0.28 (\pm 0.14) R/R_0 $ with $R_0=5.4'$. The
  almost constant metallicity appears to be consistent with a constant
  X-factor.
  
}

The $3\sigma$ limit with resolutions of $11''$ and 5\,kms$^{-1}$
corresponds to a mass of $1.7\,10^5$\,\msun. The spatial resolution of
450\,pc does not allow to detect individual GMCs if their typical size
is $\sim50$\,pc. The individual ``clumps'' delineating the spiral arms
like beads on a string (Fig.\,\ref{fig_mapco21}) have been labeled
giant molecular associations (GMAs).  These may be bound clusters of
GMCs as suggested by \citet{rand_kulkarni1990} or random
superpositions of molecular clouds and GMCs \citep{gb1993no2}.

The total mass of the molecular gas of M51 derived from the CO 2--1
data set is $1.94\,10^9$\,\msun.
%
%
%
This value agrees well with the total mass of $1.6\,10^9$\,\msun\ 
derived by \citet{helfer2003} from NRAO\,12m CO 1--0 data when using
the same CO-to-H$_2$ conversion factor and distance.

\subsection{\HI\ and radio continuum}   
\label{hi_rc_maps}
   
The large-scale distribution of the 21\,cm line of atomic hydrogen in
M51 was analyzed by \citet{rots1990} using the VLA. Note that the
total flux is in reasonable agreement with single-dish observations
\citep{rots1980}.  The \HI\ emission at $13''$ resolution
(Fig.\,\ref{fig_maphi}) is weak in the inner region while the outer CO
arms are clearly delineated in \HI.  There is patchy \HI\ emission in
the south,
{ near (0,$-250''$),}
where only very little CO is found.
%
{ In the inner arms of M51, inside the corotation radius of
  $\sim7.4$\,kpc \citep{gb1993no2}, \HI\ and H$\alpha$ emission
  \citep{rand_kulkarni1990,tilanus_allen1991,scoville2001} as well as
  young star cluster complexes \citep{bastian2005} are seen slightly
  towards the convex side, i.e.  downstream relative to the CO
  emission, suggesting that they arise when GMCs are destroyed by
  short-lived OB stars.  }
%
Figure\,\ref{fig_maphi} shows that the \HI\ clouds tracing the
south-western spiral arm, peak on the concave side of the CO arm.
This is outside the corotation radius and thus again downstream,
consistent with the above interpretation.

   
{ Figure\,\ref{fig_map20cm} shows a map of the radio continuum (RC)
  at 20\,cm (1.4\,GHz) taken with the VLA at $15''$ resolution 
  \citep{patrikeev2006}.  
  Further below, we will use this map to derive the star formation
  rate per unit area.  }
The radio continuum map shows strong emission in the inner arms, which
slowly drops off towards the outer arms.  There is patchy 20\,cm
emission in the south where there is \HI\ but only very little CO.
Again similar to \HI, the 20\,cm emission tracing the south-western
spiral arm peaks on the concave, downstream side of the CO arm. The
spiral arm near ($-40'',125''$) shows an interesting discontinuity and
local lack of radio continuum, \HI, and CO emission. 
{ \citet{patrikeev2006} discuss systematic offsets
  between the spiral arms traced by 6\,cm radio continuum emission,
  ISOCAM data at 15\,$\mu$m, and BIMA CO 1--0 emission over scales of
  several kpc.  }

\begin{figure}[htb]   
  \centering   
  \includegraphics[angle=-90,width=8cm]{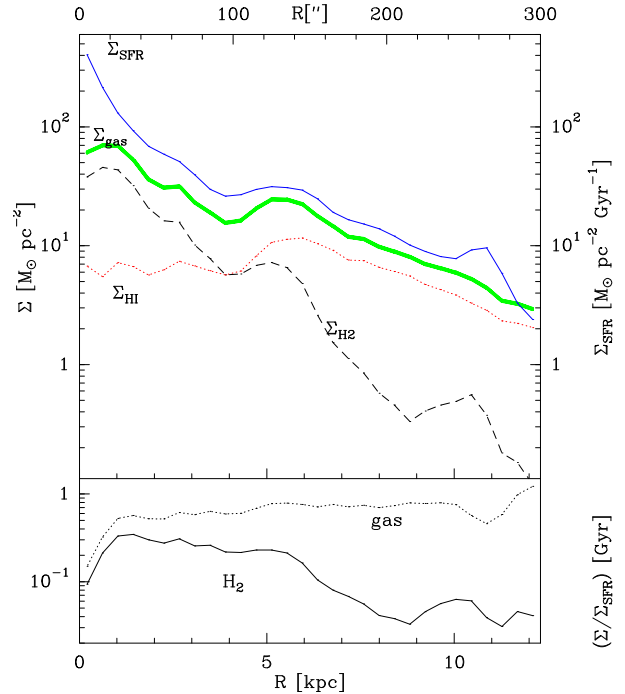}
\caption{Radial distributions of surface densities of H$_2$, \HI,    
  the total gas, and the star formation rate per unit area in M51.
  The lower box shows the ratios $\Sigma_{\rm gas}/\Sigma_{\rm SFR}$
  and $\Sigma_{{\rm H}_2}/\Sigma_{\rm SFR}$.
}
%
\label{fig_profile}   
\end{figure}   
   
\begin{figure}[h]   
\centering   
\includegraphics[angle=-90,width=8cm]{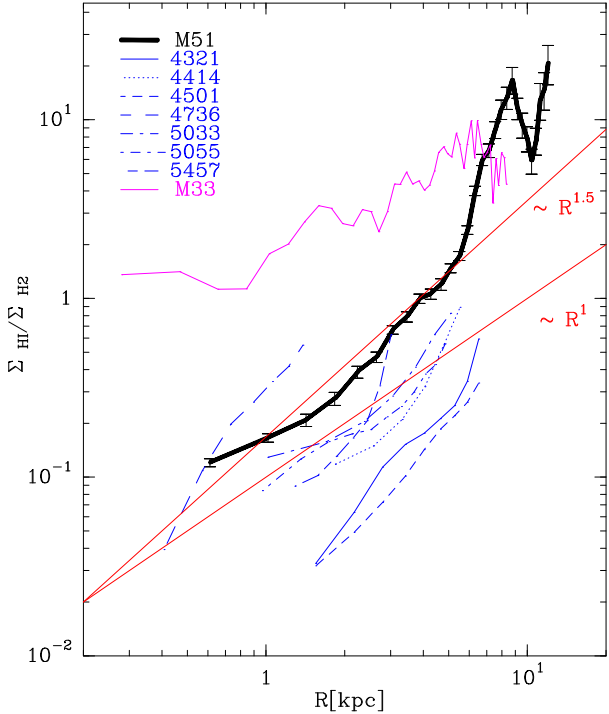}
\caption{ Radial distribution of the ratio of atomic $\sum_{\rm   
    HI}$ to molecular surface densities $\sum_{{\rm H}_2}$ for M51
  (thick line), the six spirals of the sample of
  \citet{wong_blitz2002}, and M33 \citep{heyer2004}.  For comparison,
  we show two power-laws, one where the ratio scales with $R^{1.5}$
  and one where it scales with $R$. }
\label{fig_radialhih2}   
\end{figure}   
   
\begin{figure}[h]   
\centering   
\includegraphics[angle=-90,width=8cm]{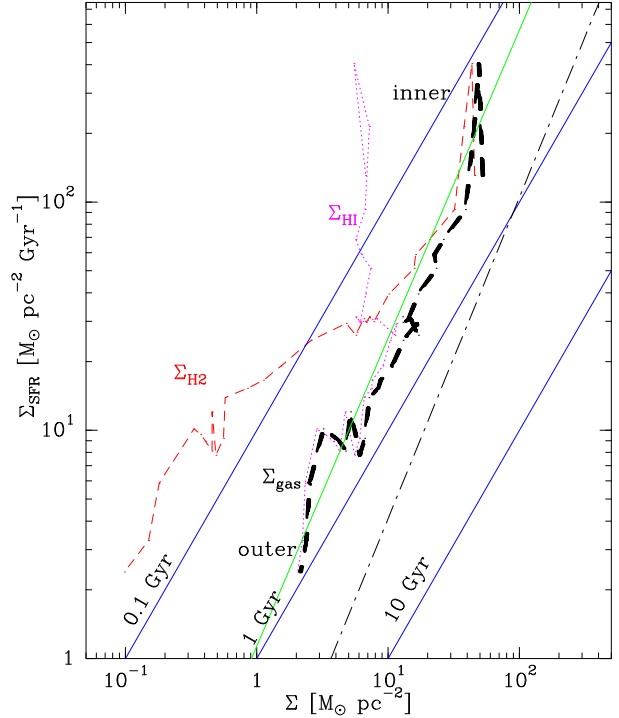}
\caption{Radially averaged star formation   
  rate per unit area, $\sum_{\rm SFR}$, versus surface densities of
  the total gas $\sum_{\rm gas}$, and of H$_2$ and \HI\ only.  The
  solid green line is the local Schmidt-law found in M51. The
  dashed-dotted black line is the global Schmidt-law found by
  \citet{kennicutt1998}.  Drawn blue lines represent lines of constant
  gas depletion time or star formation efficiency. }
\label{fig_depl}   
\end{figure}

\section{Results}   

We investigate the correlation of H$_2$ and \HI\ surface densities
with star forming activity along radial averages, including the
outskirts of the disk where \HI\ dominates.

Various tracers of star formation such as H$\alpha$, UV, the far
infrared continuum, and the radio continuum are frequently used.
H$\alpha$ emission is directly linked to OB associations but subject
to extinction. See \citet{bastian2005} for a recent study of the local
star formation rate in M51, derived from H$\alpha$ images.  Optically
thin
{ far-infrared (FIR) }
emission is usually taken as the most direct indicator of star
formation. In a recent study, \citet{calzetti2005} used MIPS/Spitzer
24\,$\mu$m maps to study the star formation rate.  Some questions
remain however 
{ in regions of diffuse FIR emission from the atomic ISM.  In such
  regions the dominant contribution to the FIR flux may be due to
  radiation heating of dust by the normal interstellar radiation field
  \citep{cox_mezger1989} and thus not trace star formation.  }

The radio continuum has widely been accepted as an alternative measure
for star formation activity, a fact which is underlined by the very
strong FIR/RC correlation in many different galaxies \citep[see the
review by][]{condon1992}. The FIR/RC correlation holds also on local
scales of a few hundred parsecs comparable to the resolution of our
data set.  See \citet[][]{murphy2005} who recently studied this
relation in M51.
Here, we use a new large scale 20\,cm map of 
\citet{patrikeev2006} 
at $15''$ resolution and assume that the 20\,cm radiation can be used
as a direct indicator for the star formation rate. We use the FIR/RC
correlation and a subsequent SFR/FIR conversion to derive the star
formation rate.  We then compare the star formation rate with the
H$_2$ and \HI\ surface densities in order to check for star formation
thresholds. We further use the measured CO 2--1 line widths to study
the gravitational stability of the gas disk.


%
%
%
 

\subsection{Radial distribution of the gas}   
   
Total H$_2$ column densities are derived from the integrated CO 2--1
intensities as described in
Sec.\,\ref{sec-molecular-gas-distribution}. We derive the face-on
surface density via $\sum_{{\rm H}_2}=2\,m_{\rm H} N({\rm H}_2) \cos{i}$
{ with the atomic hydrogen mass $m_{\rm H}$ and the inclination
  angle $i$ of M51 (Table\,\ref{tab_m51_properties}). }
The total \HI\ column density is derived assuming optically thin
emission: $N({\rm HI})=1.82\,10^{18}\,\int T_{\rm mb} dv/({\rm
  Kkms}^{-1})\,{\rm cm}^{-2}$. The corresponding surface density is:
$\sum_{\rm HI}=m_{\rm H}\,N({\rm HI})\,\cos{i}$.
{ The total molecular mass of M51 is $1.94\,10^9$\,\msun\ 
  (Sec.\,\ref{sec-molecular-gas-distribution}) and the global ratio of
  \HI\ over H$_2$ mass is 1.36. The molecular gas content of M51 is
  similar to the total molecular mass of the Milky Way,
  $1.3\,10^9$\,\msun\ \citep{misiriotis2006}. However, the Milky Way
  has a larger fraction of \HI\ mass. The ratio of \HI\ over H$_2$
  mass in the Milky Way is 6.3. 
}
   
Figure\,\ref{fig_profile} shows the radial profiles of the surface   
densities which were created by averaging in elliptical annuli spaced   
by $10''$, including points of no detection, and centered at the 0/0   
position. The annuli were assumed to be circular rings viewed at an   
inclination angle $i$ and with a line of nodes rotated from north to   
east by the position angle PA as listed in   
Table\,\ref{tab_m51_properties}.    
   
Total gas surface densities are given by $\sum_{\rm
  gas}=1.36(\sum_{{\rm H}_2}+\sum_{\rm HI})$ which includes helium.
The radially averaged $\sum_{\rm gas}$ drops by a factor of $\sim20$
from 70\,\msun\,pc$^{-2}$ in the center to 3\,\msun\,pc$^{-2}$ at
radii of 12.2\,kpc ($300''$) (Fig.\,\ref{fig_profile}). 
 

Figure\,\ref{fig_radialhih2} shows the variation of $\sum_{\rm 
  HI}/\sum_{{\rm H}_2}$ with radius in M51.  Standard deviations of 
the mean $\sum_{\rm HI}$ and $\sum_{{\rm H}_2}$ are calculated from 
the rms values along the annular averages: $\sigma = {\rm rms}\,(N_{\rm 
  pix}/n_{\rm beam})^{-0.5}$, following \citet{wong_blitz2002} with 
the number of pixels per annulus $N_{\rm pix}$ and the number of 
pixels per beam $n_{\rm beam}$. 
The fraction of atomic gas gradually increases with radius.  In the
inner regions of M51, up to $\sim6$\,kpc, the ratio of HI to H$_2$
surface density scales roughly with $R^{1.5}$.  As the radial HI
profile in M51 is roughly constant, varying only by a factor of
$\sim5$, the power-law dependence of the ratio of molecular and atomic
gas is largely due to the decrease of the molecular gas surface
density which drops by more than 2 orders of magnitude.  The \HI\ gas
surface density starts to exceed the molecular surface density at
radii greater than 4\,kpc. The outer regions show deviate from any
simple power-law.  The peak $\sum_{\rm HI}/\sum_{{\rm H}_2}$ fraction
reaches values of 20 in the outer areas of our map.
 
{ We also plot the fraction of atomic gas for the seven
  spirals studied by \citet{wong_blitz2002} and for M33 studied by
  \citet{heyer2004}.  In the center regions, the $\sum_{\rm
    HI}/\sum_{{\rm H}_2}$ ratio of the nine galaxies varies strongly
  between 0.03 in NGC\,4321 and slightly more than 1 in M33. With the
  exception of M33, the inner regions are clearly dominated by
  molecular material.  For the galaxies of the Wong sample, the ratio
  scales roughly with $R^{1.5}$ for radii upto $\sim6$\,kpc.  In M33,
  the atomic gas dominates the total gas surface density for the
  entire disk, in strong contrast to the other 8 spirals
  (Fig.\,\ref{fig_radialhih2}).  The slope is much more shallow: the
  fraction of atomic gas scales with $R^{0.6}$ only \citep{heyer2004}.
  At $\sim7$\,kpc, the fraction equals the fraction found at that
  distance in M51.
  
  Next, we compare the total gas surface densities with the star
  formation rate derived from the 20\,cm radio continuum data.  }

\subsection{Star formation rate and local Schmidt law}   
\label{sec-sfr-schmidt}

{ The radio continuum emission of the disk of M51 reflects
  the spatial distribution of the current star formation rate.
%
%
  In normal galaxies, such as M51, most of the radio continuum at
  20\,cm is non-thermal synchrotron emission \citep{condon1992},
  radiated by cosmic rays interacting with the magnetic fields of the
  interstellar medium. The cosmic rays in turn are emitted by
  supernova remnants.  The remaining radio continuum emission is
  free-free bremsstrahlung emission from thermal electrons in
  \HII-regions.  The radio continuum thus traces the current star
  formation rate of massive stars. }

The observed optically thin radio continuum emission is known to be
well correlated with FIR dust continuum emission in a large variety of
sources, including normal disk galaxies.  The tight correlation was
confirmed by \citet{helou1985} using IRAS data and is described by the
parameter
   
\begin{equation}   
  q = \log\Bigl(\frac{{\rm FIR}}   
                     {[3.75\,10^{12}\,{\rm Wm}^{-2}]}   
          \Bigr)    
      - \log\Bigl(\frac{S_{20{\rm cm}}}   
                       {\rm [Jy]}   
            \Bigr).   
\end{equation}   
   
The parameter $q$ was found to be 2.3 with a dispersion of
$\sim0.2$\,dex in a wide variety of sources \citep[][]{condon1992}. We
used this relation to derive the FIR flux at $15''$ resolution at each
position in M51.  \citet{murphy2005} have recently confirmed that this
correlation holds for a map of M51 at $70\,\mu$m using MIPS/Spitzer at
$19''$ resolution. In the disk, they find a median and scatter of
$q_{70}=1.94\pm0.19$.  Note that the correlation is expected to break
down at yet smaller scales of a few hundred parsecs around massive
star-forming regions \citep{boulanger1988}.
   
The FIR intensity reflects the current star formation rate \citep[see
e.g.][]{yun2001}, as young stars form deeply embedded in their
parental molecular clouds, before dispersing their environment by
forming \HII\ regions and by supernovae explosions. Following the
argument of \citet{thronson1986}, the FIR luminosity is proportional
to the star formation rate:
   
\begin{equation}   
L_{\rm FIR} = {\rm SFR}\, t_{\rm FIR}\, L/M    
\end{equation}   
   
where $L/M$ is the luminosity-to-mass ratio of the young stellar
clusters. Assuming a typical disruption time-scale of $t_{\rm
  FIR}=2\,10^6$\,yr and a Salpeter initial mass function (IMF), the
current star formation rate is SFR$=6.5\,10^{-10}\,(L_{\rm
  FIR}/L_{\odot})$\,\msun\,yr$^{-1}$. In summary, the star formation
rate per unit area is related to the flux density at 20\,cm via
   
\begin{equation}   
\frac{\sum_{\rm SFR}}{[{\rm M}_{\odot}\,{\rm pc}^{-2}\,{\rm Gyr}^{-1}]}   
  =   
 1.53\,10^5\,\frac{S_{20{\rm cm}}}{[{\rm Jy\,beam}^{-1}]}.   
\end{equation}   

{ The gas depletion or consumption time is defined as
  $\tau_{gas}=\sum_{\rm gas}/\sum_{\rm SFR}$. This is the time which
  would be needed to convert the total gas content into stars assuming
  that the SFR is constant with time and that there is no gas infall
  or recycling via stellar winds. Note that the inverse of the gas
  depletion time is the star formation efficiency.
  
  The global star formation rate of M51 is 2.56\,\msun\,yr$^{-1}$ and
  the global gas depletion time 0.76\,Gyr.  The latter value
  corresponds to a global star formation efficiency (SFE) of 13\% per
  0.1\,Gyr.  Both values, the depletion time and the SFE, agree to
  within a factor of 2 with the values derived by \citet{scoville2001}
  from Hubble Space Telescope (HST) H$\alpha$ images and CO data.
  \citet{misiriotis2006} have recently studied the distribution of the
  ISM in the Milky Way.  They find a similar SFR of
  2.7\,\msun\,yr$^{-1}$ but factor $\sim5$ larger gas depletion time
  of 3.57\,Gyr in the Milky Way. Typical normal spiral galaxies need
  less time than found in the Milky Way, but more time than found in
  M51, to consume all the gas into stars.  \citet{kennicutt1998} finds
  a median depletion time of 2.1\,Gyr for his sample of 61 normal disk
  galaxies which shows variations between 0.2\,Gyr in starburst
  galaxies like NGC\,5169 and 12\,Gyr in early-type spirals such as
  M31.
}

{ The SFR peaks above 100\,M$_{\odot}$pc$^{-2}$Gyr$^{-1}$
  in the center (Fig.\,\ref{fig_profile}), indicating a nuclear
  starburst, and drops radially to values of
  3\,M$_{\odot}$pc$^{-2}$Gyr$^{-1}$ at 12\,kpc distance.  Note that
  the C2 cluster complex studied by \citet{bastian2005} shows a local
  SFR of 2600\,M$_{\odot}$pc$^{-2}$Gyr$^{-1}$, while the other
  complexes studied by these authors show moderate local rates of
  $\sim60-70$\,M$_{\odot}$pc$^{-2}$Gyr$^{-1}$. Regions of more than
  $\sim100\,$M$_{\odot}$pc$^{-2}$Gyr$^{-1}$ (i.e.
  0.1\,M$_{\odot}$kpc$^{-2}$yr$^{-1}$) are classified as starbursts
  \citep[e.g.][]{kennicutt1998}.
  The center of M51 harbours a Seyfert 2 AGN surrounded by a $\sim100$\,pc
  disk/torus \citep{kohno1996} of warm and dense gas
  \citep{matsushita1998,Matsushita2004}. }

Figure\,\ref{fig_profile} shows the variation of the radially averaged
star formation rate in comparison with the \HI, H$_2$, and total gas
surface densities in M51.
%
Ignoring the center, the radial drop of the star formation rate
closely resembles the drop of the total gas surface density. Neither
does the SFR show the rather flat distribution of the atomic gas nor
the much steeper drop of the molecular gas.
{ In contrast, we would have anticipated a good correlation
  of the star formation rate with the molecular gas since star
  formation is known to occur only in molecular clouds. Indeed,
  \citet{heyer2004} find a strong correlation between the star
  formation rate and the molecular gas surface density in M33.
  Similarly, \citet{wong_blitz2002} report a much better correlation
  of $\Sigma_{\rm SFR}$ with $\Sigma_{\rm H2}$ than with $\Sigma_{\rm
    HI}$ in their sample of 6 molecule-rich spirals. However,
  \citet{kennicutt1998} who studied 88 galaxies and found that the
  disk-averaged SFR is much better correlated with the disk-averaged
  \HI\ surface densities than with the H$_2$ surface densities. In
  M51, \HI\ is often found downstream of the CO arms indicating that
  the \HI\ clouds are the remnants of GMCs photo-dissociated by young
  massive stars (Sec.\,\ref{hi_rc_maps}). The star formation rate may
  thus regulate the surface density of the atomic gas and hence
  explain the observed correlation. }

Figure\,\ref{fig_depl} shows the variation of SFR in M51 with gas
surface densities together with lines of constant gas depletion time.
The gas consumption time in M51 varies between 0.1\,Gyr in the center
where the gas surface density and the SFR surface density are high and
1\,Gyr at large radii where the gas surface density and SFR density
are low (Figs.\,\ref{fig_profile},\ref{fig_depl}).  In contrast, the
Milky Way \citep{misiriotis2006} and the 6 CO-bright spiral galaxies
studied by \citet{wong_blitz2002} exhibit gas depletion times which
are larger by a factor of about 10. They rise from $\sim1$\,Gyr in the
centers to $\sim10$\,Gyr and slightly more in the outskirts.
 
Figure\,\ref{fig_depl} also shows that the star formation rate is 
proportional to a power of the total gas surface density im M51 
 
\begin{equation} 
  {\sum}_{\rm SFR} = A {{\sum}^n_{\rm gas}} 
\label{eq-schmidt-m51}
\end{equation} 
 
i.e. follows a Schmidt-law \citep{schmidt1959}, with 
{ normalization
}
$A=1.1\pm0.5$ and the slope $n=1.4\pm0.6$
where ${\sum}_{\rm SFR}$ is in units of \msun pc$^{-2}$ Gyr$^{-1}$ 
and ${\sum}_{\rm gas}$ in units of \msun pc$^{-2}$. 
 
\begin{center} 
\begin{table*} 
\begin{tabular}{lrrrrr} 
\hline \hline 
 & $A$              & $n$                & Reference \\ 
\hline 
{\bf Local Schmidt-laws:} \\
M51                              & $1.12\pm 0.50$    & $1.35\pm 0.61$ & this paper             \\ 
M33                              & $3.5\pm66$        & $3.3\pm 0.07$  & \citet{heyer2004}      \\ 
Milky Way                        &                   & $2.18\pm0.20$  & \citet{misiriotis2006} \\
7 CO-bright spiral galaxies      &                   & $1.7\pm0.3$    & \citet{wong_blitz2002} \\ 
16 spiral galaxies               &                   & $\sim2$        & \citet{boissier2003}   \\
simulations                      & $0.25\pm 0.16$    & $1.31\pm 0.15$ & \citet{li2006}         \\ 
{\bf Global Schmidt-laws:} \\
97 normal and starburst galaxies & $0.25\pm 0.07$    & $1.4\pm 0.15$  & \citet{kennicutt1998}  \\ 
7 CO-bright spiral galaxies      &                   & 1.7            & \citet{wong_blitz2002} \\ 
simulations                      & $0.11\pm0.04$     & $1.56\pm0.09$  & \citet{li2006}         \\ 
\hline 
\end{tabular} 
\label{tab_schmidt} 
\caption[]{\label{tab_schmidt}   
{ Schmidt Law ${\sum}_{\rm SFR} = A {{\sum}^n_{\rm gas}}$
observed in M51 and other galaxies and samples of galaxies in
comparison with the Schmidth-Law derived simulations. ${\sum}_{\rm
SFR}$ and $A$ are in units of \msun pc$^{-2}$ Gyr$^{-1}$ and ${\sum}_{\rm gas}$
in units of \msun pc$^{-2}$.  The values of the sample of
\citet{wong_blitz2002} hold for an extinction correction that depends
on the gas column density. }
%
}   
\end{table*} 
\end{center}  


  In Table\,\ref{tab_schmidt}, we compare the slope found
  in M51 with the results obtained for other galaxies and the results
  from simulations. The slope found in M51 agrees with the global
  Schmidt-law, seen in a study of disk-averaged ${\sum}_{\rm gas}$ and
  ${\sum}_{\rm SFR}$ of 61 normal and 36 starburst galaxies
  \citep{kennicutt1998}. Note however that the slope of the 61 normal
  galaxies is much less well-defined. Depending on the fitting method
  it varies between 1.3 and 2.5 \citep{kennicutt1998}.
  
  In contrast, the observed slopes of {\it local} Schmidt laws,
  describing radial averages of $\Sigma_{\rm gas}$ and $\Sigma_{\rm
    SFR}$ in individual galaxies, do not in general agree with the
  global value, but vary strongly between 1.2 and 3.3.
  \citet{wong_blitz2002} studied radial averages of ${\sum}_{\rm gas}$
  and ${\sum}_{\rm SFR}$ in a sample of spiral galaxies and derive
  local Schmidt laws with slopes between 1.2 and 2.1, assuming that
  extinction depends on gas column density.  \citet{boissier2003}
  study 16 spiral galaxies to study the local star formation laws and
  find a slope of $\sim2$. In a similar study, \citet{heyer2004} shows
  that M33 exhibits a significantly steeper slope of $3.3$.
  \citet{misiriotis2006} find a slope of $2.2$ for the Milky Way.  

\subsection{Gravitational stability}   

{ Exceeding a critical gas density may lead to the
  formation of clouds and possibly stars. The critical density may in
  turn be determined by gravitation.
}
Analogous to the well known Jeans stability criterion, the influence
of differential rotation of a homogeneous, thin disk can be described
by a dispersion relation for axially symetric disturbences
\citep{kley_peitz2004}: $\sigma^2=\kappa^2 + \sigma_{\rm gas}^2k^2 -
2\pi G \sum |k|$. Here, $\sigma$ is the oscillation frequency,
$\sigma_{\rm gas}$ the gas velocity dispersion, $k$ the wavenumber,
and $\kappa$ the epicyclic frequency
$\kappa^2=\frac{2V}{R}\left(\frac{V}{R}+\frac{dV}{dR}\right)$.  For a
flat rotation curve, $\kappa^2=2\Omega^2$.  The system becomes
unstable when the Toomre parameter \citep{toomre1964}
\begin{equation}  
Q_{\rm gas}(R) \equiv \frac{\Sigma_{\rm crit}(R)}{\Sigma_{\rm gas}(R)}
       = \frac{\kappa(R) \sigma_{\rm gas}(R)}{\pi G \Sigma_{\rm gas}(R)},  
\label{eq_sigma_crit}
\end{equation}  
describing the competition of pressure and rotation on the one side
and gravitation on the other side, at a given radius, is less than 1,
i.e. when the gas surface density exceeds the critical surface
density.  

{ In the following, we will assume that the disk of M51 is
  almost unstable as predicted by disk models
  \citep[e.g.][]{lin_pringle1987}, i.e. $Q_{\rm gas}=1$, to calculate the
  critical gas velocity dispersion $\sigma_{\rm crit}$ necessary to
  stabilize the gas against gravitational collapse. We will then
  compare $\sigma_{\rm crit}$ with the observed velocity dispersion of
  the molecular gas, $\sigma_{\rm CO}$, as function of the
  galacto-centric radius.

}

\subsubsection{Rotation curve}

The rotation curve of M51, needed to calculate the epicyclic frequency
and the critical gas density (Eq.\,\ref{eq_sigma_crit}), was derived
by \citet{gb1993no1,gb1993no2} from the CO 2--1 position-velocity
diagram along the major axis of M51, using the velocities at the peak
intensities, and correcting for the inclination. The rotation curve
(Fig.\,\ref{fig_radial_vel_disp}) rises steeply within $10''$ of the
center indicating the presence of a compact nuclear mass component in
addition to the central bulge and then stays constant at
$\sim200$\,kms$^{-1}$ out to a radius of 10\,kpc. We assume here that
the rotation curve stays almost constant further out to at least
12\,kpc, ignoring any effect of the companion galaxy.  The definition
of the rotation curve used here appears to be more appropriate for the
face-on galaxy M51 than the definition used by \citet{sofue1996} who
derived the rotation curve from the terminal velocities at which the
peak intensity has dropped to 20\% of its value, which leads to very
high velocities of upto 260\,kms$^{-1}$ at radii between 3 and 9\,kpc.


\subsubsection{Critical velocity dispersion}

{

Figure\,\ref{fig_radial_vel_disp} also shows the radially averaged
critical velocity dispersion $\sigma_{\rm crit}$ derived from the gas
surface density and the rotation curve, assuming $Q_{\rm gas}=1$
(Eq.\,\ref{eq_sigma_crit}). The critical dispersion, necessary to
stabilize the gas against gravitational collapse, lies between 1.7 and
6.8\,kms$^{-1}$. The dispersion peaks near $R=5.5$\,kpc where the H$_2$
surface density is high (Fig.\,\ref{fig_profile}). It then declines
slowly to $\sim2$\,kms$^{-1}$ at $R=12$\,kpc.

}

\subsubsection{Observed velocity dispersion}

\begin{figure}[h]   
   \centering  
   \includegraphics[angle=-90,width=8cm]{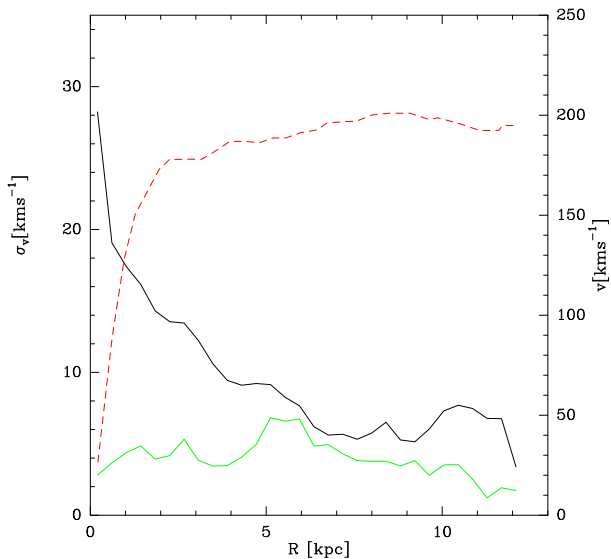}
 \caption{{\bf Left:} The black drawn line shows the 
   $^{12}$CO 2--1 gas dispersion $\sigma_{\rm CO}$ derived from the
   observed equivalent widths. Positions without detection of
   $^{12}$CO are ignored.
   { The green drawn line shows the gas dispersion $\sigma_{\rm crit}$,
     assuming gravitational stability, i.e. $\Sigma_{\rm
       crit}=\Sigma_{\rm gas}$.  }
{\bf Right:} The dashed line gives the rotation curve derived by
\citet{gb1993no2}. }
 \label{fig_radial_vel_disp}   
 \end{figure}

{ The velocity dispersion of the molecular gas is estimated
  here from the equivalent $^{12}$CO 2--1 line widths $\Delta v_{\rm
    eq}=\int T dv / T_{\rm pk}$ via $\sigma_{\rm CO}
  = \Delta v_{\rm eq}/(2\,\sqrt{2\,\ln2})$.
  Figure\,\ref{fig_radial_vel_disp} shows the radially averaged
  velocity dispersion $\sigma_{\rm CO}$.  It drops from
  $\sim28$\,kms$^{-1}$ in the center to $\sim6$\,kms$^{-1}$ at radii
  of 7 to 9\,kpc.  Further out, it rises again to values of
  $\sim8\,$kms$^{-1}$. The rise of the observed dispersion at radii of
  10-13\,kpc, is due to the increased line widths in the companion
  galaxy.
  
  Similar vertical velocity dispersions are derived from \HI\ 
  observations of spiral galaxies. Typically, these vary radially from
  $\sim12-15$\,kms$^{-1}$ in the central parts to
  $\sim4-6$\,kms$^{-1}$ in the outer parts \citep[see review
  of][]{dib2006}.
  
  
  The CO velocity dispersion observed in M51 exceeds the critical
  dispersion at all radii. The ratio between the observed and the
  critical dispersions drop from a factor of $\sim5$ to almost 1 for
  radii between 1 and 5\,kpc. This may indicate that the gas is
  stabilized against collapse at all radii.
  
  However, the observed gas dispersions are averages over the 450\,pc
  beam and along the elliptical annuli, and are broadened relative to
  the intrinsic dispersions due to systematic motions which do not
  contribute necessarily to local support of the gas against gravity.
  Moreover, the influence of stars is neglected.
   }



\section{Discussion} 
\label{sec-discussion}

%
{

\subsection{Toomre stability}
  
Several authors have argued that galactic disks are self-regulated
through gravitational instabilities to have their Toomre $Q$
parameter of the order of 1 \citep[see e.g.][]{combes2001}. As soon as
gas dissipation leads to $\Sigma_{\rm gas}>\Sigma_{\rm crit}$ at a
given radius, the disk becomes gravitationally unstable, leading to
density waves transferring angular momentum, causing dissipation and
heating of the gas and thus an increase of the velocity dispersion,
stabilizing the disk again.  Models indeed indicate that the disk
settles at the border of instability \citep{lin_pringle1987}.
  
However, in general the velocity dispersion and surface density of
stars may play an important role in the stability analysis.
\citet{bottema1993} discuss the dispersion of stellar disks and the
self-regulating mechanism that keeps $Q$ near 1 over the entire disk
of spiral galaxies. To take this into account, a total $Q$ parameter
has been used by e.g. \citet{combes2001}: $Q_{\rm tot}^{-1} = Q_{\rm
  gas}^{-1} + Q_{\rm star}^{-1}$. Neglecting stars is valid only when
the gas surface density divided by the gas dispersion dominates all
other $\Sigma/\sigma$ terms.

Observationally, $Q_{\rm gas}$ has often been found to be larger than
1 by factors of a few.  \citet{kennicutt1989} derived radial averages
of the Toomre-parameter of a sample of 15 galaxies, assuming a
constant velocity dispersion of 6\,kms$^{-1}$, and found that star
formation is restricted to regions where $Q_{\rm gas}<1.5$.  A similar
conclusion was drawn from a study of 32 nearby spiral galaxies by
\citet{martin_kennicutt2001} with few exceptions.\footnote{Both,
  \citet{kennicutt1989} and \citet{martin_kennicutt2001} included M51
  in their sample, using FCRAO $50''$ CO 1--0 data combined with
  H$\alpha$ data. However, the Toomre parameter and its variation with
  radius in M51 are not explicitly discussed. }  Similarly,
\citet{wong_blitz2002} also assume a constant velocity dispersion and
find $\Sigma_{\rm gas}\approx\Sigma_{\rm crit}$ over a wider range of
radii for some of their galaxies while others show deviations of
factors of $\sim2$.

While in the inner 4\,kpc of M51, the gas dispersion is large and
$Q_{\rm gas}$ lies between 2 and 5, the stars may contribute here, and
make the disk marginally stable. Streaming motions are a consequence
of spiral waves and gravitational instabilities \citep{combes2001}.
M51 shows strong streaming motions of $60-150$\,kms$^{-1}$ in the
plane of the galaxy in the inner spiral arms as can be seen in the
interferometric CO maps of \citet{aalto1999}. The broadening of the
velocity dispersions due to streaming motions is naturally taken into
account here.  However, the observed dispersion is also broadened by
the very steep, unresolved rotation curve in the center of M51. A more
careful stability analysis would need to correct the observed
dispersions for rotation gradients, take into account the influence of
stars, and study individual regions of M51 at high spatial resolution.

\subsection{The local Schmidt law and star formation efficiency}

Recent smoothed particle hydrodynamics (SPH) simulations of
gravitational instability of isolated disk galaxies comprising a dark
matter halo, a disk of stars, and isothermal gas by
\citet{li2005,li2006} find a slope of 1.31 of the local Schmidt law
(Table\,\ref{tab_schmidt}), close to the slope of the global Schmidt
law, which is also reproduced, and close to the slope of 1.35 found in
M51. On the other hand, the slope of 1.3 is at the low-end of the
local Schmidt law slopes observed so far. 

We find a normalization factor $A$ of the local Schmidt law in M51 of
1.12\msun pc$^{-2}$ Gyr$^{-1}$ (Table\,\ref{tab_schmidt}), similar to
$A$ of gas-rich models in \citet{li2006}. This is a factor 5 larger
than the average value of the \citet{li2006} models. It is also a
factor 5 larger than the normalization found in the large sample of
\citet{kennicutt1998}. 

Interestingly, the normal galaxies of the \citet{kennicutt1998} sample
show on average a factor $\sim3$ longer gas consumption times than
found in M51. In Section\,\ref{sec-sfr-schmidt} we also noticed that
the Milky Way \citep{misiriotis2006} and the 6 CO-bright galaxies
studied by \citet{wong_blitz2002} exhibit consumption times which are
larger than the consumption time found in M51 by about one order of
magnitude. Apparently, the star formation efficiency of M51 is higher
than in many other normal galaxies.

The reason may be the interaction of M51 with its neighbouring galaxy
NGC\,5195, as has been proposed before as mentioned in the
Introduction
\citep[e.g.][]{nikola2001,howard1990,toomre_toomre1972}. This is
supported by the recent simulations of \citet{li2004} which show that
interaction and merging can lead to instable disks and strong
starbursts.

%


}

\section{Summary} 
  
We mapped completely the interacting spiral galaxy M51 in the
$^{12}$CO 2--1 line using HERA at the IRAM-30m telescope. The map
includes the companion galaxy as well as the south-western arm out to
radii of 12\,kpc at linear resolutions of 450\,pc ($11''$). These data
were combined with maps of \HI\ and the radio continuum at 20\,cm at
similar resolutions to study radially averaged surface densities. The
star formation rate per unit area was estimated from the radio
continuum, allowing to study star formation laws like the Schmidt law.
The critical density for gravitational instability was compared with
the total gas surface density along the radial averages. In detail, we
obtain the following results:

\begin{itemize}  
  
\item { Global properties: The total H$_2$ mass of M51 is
    $1.94\,10^9$\,\msun\ and
    the global \HI/H$_2$ mass ratio is 1.36. The global star formation rate
    is 2.56\,\msun\,yr$^{-1}$ and the global gas depletion time is
    $0.8$\,Gyr.}

\item The total gas surface density drops from
  $\sim$70\,\msun\,pc$^{-2}$ in the center to
  $\sim3$\,\msun\,pc$^{-2}$ at radii of 12\,kpc. The ratio of atomic
  to molecular gas surface density rises from 0.1 in the center to 20
  in the outer regions. It is 1 at $\sim4\,$kpc. Up to $\sim6$\,kpc
  the ratio follows a simple power law of $R^{1.5}$ similar to that
  found in other spirals. At larger radii, the simple power law
  relation breaks down.
  
  
\item { The gas depletion times in M51, vary between 0.1
    and 1\,Gyrs, and are shorter than in many other non-interacting
    normal galaxies. Simulations indeed show that interaction can lead
    to high star formation efficiencies.
    
  \item The star formation rate per unit area drops from $\sim400$
    \,\msun\,pc$^{-2}$\,Gyr$^{-1}$ in the starburst center to
    $\sim2$\,\msun\,pc$^{-2}$\,Gyr$^{-1}$ in the outskirts. It is much
    better correlated with the total gas surface density than with the
    surface densities of the molecular or atomic gas. The correlation
    follows a Schmidt law $\Sigma_{\rm SFR}\propto\Sigma_{\rm gas}^n$
    with an index of $n=1.4\pm0.6$. Only few studies of local Schmidt
    laws exist to date.  The slopes of local Schmidt laws observed in
    other spirals vary strongly between 1.2 and 3.3.  The slope of 1.4
    agrees with recent SPH simulations of isolated disk galaxies
    \citet{li2005}.  }
  
  
\item The radially averaged dispersion of the observed $^{12}$CO 2--1
  lines varies between 28\,kms$^{-1}$ in the center and
  $\sim$6\,kms$^{-1}$ in the outskirts.
  
\item { We investigated the gravitational stability using
    the Toomre criterion.  The critical gas velocity dispersions
    needed to stabilize the gas against gravitational collapse in the
    differentially rotating disk of M51, vary with radius between 1.7
    and 6.8\,kms$^{-1}$. Observed radially averaged dispersions
    derived from the CO data exceed the critical dispersions by
    factors $Q_{\rm gas}$ of 1 to 5. Unresolved gradients of the
    rotation curve lead to an overestimate of the intrinsic velocity
    dispersion.  The gravitational potential of stars not considered
    here, may lead to a marginally stable disk.
%
}
  
\end{itemize}  
  
In a forthcoming paper we will discuss the distribution of molecular
gas, local Schmidt laws, and local critical gas surface densities,
discerning the conditions in different parts of M51, i.e. the arm,
interarm, and central regions. We will also study in more detail the
spatial and kinematic distribution of atomic and molecular clouds,
especially in the outer spiral arms.

\begin{acknowledgements}   
  We thank R.\,Beck for providing us the radio-continuum data and
  A.H.\,Rots for the HI data. We would also like to thank R.\,Beck,
  T.\,Wong, Y.\,Sofue, N.\,Scoville, and F.\,Combes for valuable
  comments and discussions.  Insightful comments from an anonymous
  referee are appreciated. We are grateful to the IRAM staff at Pico
  Veleta for excellent support at the telescope. IRAM is supported by
  INSU/CNRS (France), MPG (Germany), and IGN (Spain).
\end{acknowledgements}   
   
\bibliographystyle{aa} 
\bibliography{aamnem99,5579} 

\begin{thebibliography}{73}
\expandafter\ifx\csname natexlab\endcsname\relax\def\natexlab#1{#1}\fi

\bibitem[{{Aalto} {et~al.}(1999){Aalto}, {Huettemeister}, {Scoville}, \&
  {Thaddeus}}]{aalto1999}
{Aalto}, S., {Huettemeister}, S., {Scoville}, N.~Z., \& {Thaddeus}, P. 1999,
  ApJ, 522, 165

\bibitem[{{Adler} {et~al.}(1992){Adler}, {Lo}, {Wright}, {Rydbeck}, {Plante},
  \& {Allen}}]{adler1992}
{Adler}, D.~S., {Lo}, K.~Y., {Wright}, M.~C.~H., {et~al.} 1992, \apj, 392, 497

\bibitem[{{Arimoto} {et~al.}(1996){Arimoto}, {Sofue}, \&
  {Tsujimoto}}]{arimoto1996}
{Arimoto}, N., {Sofue}, Y., \& {Tsujimoto}, T. 1996, \pasj, 48, 275

\bibitem[{{Bastian} {et~al.}(2005){Bastian}, {Gieles}, {Efremov}, \&
  {Lamers}}]{bastian2005}
{Bastian}, N., {Gieles}, M., {Efremov}, Y.~N., \& {Lamers}, H.~J.~G.~L.~M.
  2005, \aap, 443, 79

\bibitem[{{Bell} {et~al.}(2006){Bell}, {Roueff}, {Viti}, \&
  {Williams}}]{bell2006}
{Bell}, T.~A., {Roueff}, E., {Viti}, S., \& {Williams}, D.~A. 2006, \mnras,
  371, 1865

\bibitem[{{Boissier} {et~al.}(2003){Boissier}, {Prantzos}, {Boselli}, \&
  {Gavazzi}}]{boissier2003}
{Boissier}, S., {Prantzos}, N., {Boselli}, A., \& {Gavazzi}, G. 2003, \mnras,
  346, 1215

\bibitem[{{Bottema}(1993)}]{bottema1993}
{Bottema}, R. 1993, \aap, 275, 16

\bibitem[{{Boulanger} \& {Perault}(1988)}]{boulanger1988}
{Boulanger}, F. \& {Perault}, M. 1988, \apj, 330, 964

\bibitem[{{Bresolin} {et~al.}(2004){Bresolin}, {Garnett}, \&
  {Kennicutt}}]{bresolin2004}
{Bresolin}, F., {Garnett}, D.~R., \& {Kennicutt}, R.~C. 2004, \apj, 615, 228

\bibitem[{{Calzetti} {et~al.}(2005){Calzetti}, {Kennicutt}, {Bianchi},
  {Thilker}, {Dale}, {Engelbracht}, {Leitherer}, {Meyer}, {Sosey}, {Mutchler},
  {Regan}, {Thornley}, {Armus}, {Bendo}, {Boissier}, {Boselli}, {Draine},
  {Gordon}, {Helou}, {Hollenbach}, {Kewley}, {Madore}, {Martin}, {Murphy},
  {Rieke}, {Rieke}, {Roussel}, {Sheth}, {Smith}, {Walter}, {White}, {Yi},
  {Scoville}, {Polletta}, \& {Lindler}}]{calzetti2005}
{Calzetti}, D., {Kennicutt}, R.~C., {Bianchi}, L., {et~al.} 2005, \apj, 633,
  871

\bibitem[{{Combes}(2001)}]{combes2001}
{Combes}, F. 2001, in ASP Conf. Ser. 249: The Central Kiloparsec of Starbursts
  and AGN: The La Palma Connection, ed. J.~H. {Knapen}, J.~E. {Beckman},
  I.~{Shlosman}, \& T.~J. {Mahoney}, 475

\bibitem[{{Condon}(1992)}]{condon1992}
{Condon}, J.~J. 1992, \araa, 30, 575

\bibitem[{{Cox} \& {Mezger}(1989)}]{cox_mezger1989}
{Cox}, P. \& {Mezger}, P.~G. 1989, \aapr, 1, 49

\bibitem[{{Daigle} {et~al.}(2006){Daigle}, {Carignan}, {Amram}, {Hernandez},
  {Chemin}, {Balkowski}, \& {Kennicutt}}]{daigle2006}
{Daigle}, O., {Carignan}, C., {Amram}, P., {et~al.} 2006, \mnras, 367, 469

\bibitem[{de~Vaucouleurs {et~al.}(1991)de~Vaucouleurs, de~Vaucouleurs, Corwin,
  Buta, Pasturel, \& Fouque}]{devaucouleurs1991}
de~Vaucouleurs, G., de~Vaucouleurs, A., Corwin, H.~J., {et~al.} 1991, Third
  reference catalogue of bright galaxies (New York: Springer-Verlag)

\bibitem[{{Dib} {et~al.}(2006){Dib}, {Bell}, \& {Burkert}}]{dib2006}
{Dib}, S., {Bell}, E., \& {Burkert}, A. 2006, \apj, 638, 797

\bibitem[{{Elmegreen}(1994)}]{elmegreen1994}
{Elmegreen}, B.~G. 1994, \apjl, 425, L73

\bibitem[{{Feldmeier} {et~al.}(1997){Feldmeier}, {Ciardullo}, \&
  {Jacoby}}]{feldmeier1997}
{Feldmeier}, J.~J., {Ciardullo}, R., \& {Jacoby}, G.~H. 1997, \apj, 479, 231

\bibitem[{Garcia-Burillo {et~al.}(1993{\natexlab{a}})Garcia-Burillo, Combes, \&
  Gerin}]{gb1993no2}
Garcia-Burillo, S., Combes, F., \& Gerin, M. 1993{\natexlab{a}}, A\&A, 274, 148

\bibitem[{Garcia-Burillo {et~al.}(1993{\natexlab{b}})Garcia-Burillo, Guelin, \&
  Cernicharo}]{gb1993no1}
Garcia-Burillo, S., Guelin, M., \& Cernicharo, J. 1993{\natexlab{b}}, A\&A,
  274, 123

\bibitem[{Gerin \& Phillips(2000)}]{gerin_phillips2000}
Gerin, M. \& Phillips, T. 2000, ApJ, 537, 644

\bibitem[{{Greve} {et~al.}(1998){Greve}, {Kramer}, \& {Wild}}]{greve1998}
{Greve}, A., {Kramer}, C., \& {Wild}, W. 1998, \aaps, 133, 271

\bibitem[{Gu\'elin {et~al.}(1995)Gu\'elin, Zylka, Mezger, Haslam, \&
  Kreysa}]{guelin1995}
Gu\'elin, M., Zylka, R., Mezger, P.~G., Haslam, C. G.~T., \& Kreysa, E. 1995,
  A\&A, 298, 29

\bibitem[{{Helfer} {et~al.}(2003){Helfer}, {Thornley}, {Regan}, {Wong},
  {Sheth}, {Vogel}, {Blitz}, \& {Bock}}]{helfer2003}
{Helfer}, T.~T., {Thornley}, M.~D., {Regan}, M.~W., {et~al.} 2003, \apjs, 145,
  259

\bibitem[{{Helou} {et~al.}(1985){Helou}, {Soifer}, \&
  {Rowan-Robinson}}]{helou1985}
{Helou}, G., {Soifer}, B.~T., \& {Rowan-Robinson}, M. 1985, \apjl, 298, L7

\bibitem[{{Heyer} {et~al.}(2004){Heyer}, {Corbelli}, {Schneider}, \&
  {Young}}]{heyer2004}
{Heyer}, M.~H., {Corbelli}, E., {Schneider}, S.~E., \& {Young}, J.~S. 2004,
  \apj, 602, 723

\bibitem[{{Howard} \& {Byrd}(1990)}]{howard1990}
{Howard}, S. \& {Byrd}, G.~G. 1990, \aj, 99, 1798

\bibitem[{Israel \& Baas(2002)}]{israel_baas2002}
Israel, F.~P. \& Baas, F. 2002, A\&A, 383, 82

\bibitem[{{Israel} {et~al.}(2006){Israel}, {Tilanus}, \& {Baas}}]{israel2006}
{Israel}, F.~P., {Tilanus}, R.~P.~J., \& {Baas}, F. 2006, \aap, 445, 907

\bibitem[{{Kennicutt}(1989)}]{kennicutt1989}
{Kennicutt}, R.~C. 1989, \apj, 344, 685

\bibitem[{{Kennicutt}(1998)}]{kennicutt1998}
{Kennicutt}, R.~C. 1998, \apj, 498, 541

\bibitem[{Kley(2004)}]{kley_peitz2004}
Kley, P. 2004, Theoretische Astrophysik, Wintersemester 2004/05 (Universitaet
  Tuebingen)

\bibitem[{Kohno {et~al.}(1996)Kohno, Kawabe, Tosaki, \& Okumura}]{kohno1996}
Kohno, K., Kawabe, R., Tosaki, T., \& Okumura, S. 1996, ApJ, 461, 29

\bibitem[{{Kramer} {et~al.}(2005){Kramer}, {Mookerjea}, {Bayet},
  {Garcia-Burillo}, {Gerin}, {Israel}, {Stutzki}, \& {Wouterloot}}]{kramer2005}
{Kramer}, C., {Mookerjea}, B., {Bayet}, E., {et~al.} 2005, \aap, 441, 961

\bibitem[{{Kuno} \& {Nakai}(1997)}]{kuno1997}
{Kuno}, N. \& {Nakai}, N. 1997, \pasj, 49, 279

\bibitem[{{Kuno} {et~al.}(1995){Kuno}, {Nakai}, {Handa}, \& {Sofue}}]{kuno1995}
{Kuno}, N., {Nakai}, N., {Handa}, T., \& {Sofue}, Y. 1995, \pasj, 47, 745

\bibitem[{{Li} {et~al.}(2004){Li}, {Mac Low}, \& {Klessen}}]{li2004}
{Li}, Y., {Mac Low}, M.-M., \& {Klessen}, R.~S. 2004, \apjl, 614, L29

\bibitem[{{Li} {et~al.}(2005){Li}, {Mac Low}, \& {Klessen}}]{li2005}
{Li}, Y., {Mac Low}, M.-M., \& {Klessen}, R.~S. 2005, \apjl, 620, L19

\bibitem[{{Li} {et~al.}(2006){Li}, {Mac Low}, \& {Klessen}}]{li2006}
{Li}, Y., {Mac Low}, M.-M., \& {Klessen}, R.~S. 2006, \apj, 639, 879

\bibitem[{{Lin} \& {Pringle}(1987)}]{lin_pringle1987}
{Lin}, D.~N.~C. \& {Pringle}, J.~E. 1987, \mnras, 225, 607

\bibitem[{Lord \& Young(1990)}]{lord_young1990}
Lord, S. \& Young, J. 1990, ApJ, 356, 135

\bibitem[{{Martin} \& {Kennicutt}(2001)}]{martin_kennicutt2001}
{Martin}, C.~L. \& {Kennicutt}, R.~C. 2001, \apj, 555, 301

\bibitem[{Matsushita {et~al.}(1998)Matsushita, Kohno, Vila-Vilaro, Tosaki, \&
  Kawabe}]{matsushita1998}
Matsushita, S., Kohno, K., Vila-Vilaro, B., Tosaki, T., \& Kawabe, R. 1998,
  ApJ, 495, 267

\bibitem[{Matsushita {et~al.}(2004)Matsushita, Sakamoto, Kuo, Hsieh, Trung,
  Mao, Iono, Peck, Wiedner, Liu, Ohashi, \& Lim}]{Matsushita2004}
Matsushita, S., Sakamoto, K., Kuo, C.-Y., {et~al.} 2004, ApJ, 616, 55

\bibitem[{{Meijerink} {et~al.}(2005){Meijerink}, {Tilanus}, {Dullemond},
  {Israel}, \& {van der Werf}}]{meijerink2005}
{Meijerink}, R., {Tilanus}, R.~P.~J., {Dullemond}, C.~P., {Israel}, F.~P., \&
  {van der Werf}, P.~P. 2005, \aap, 430, 427

\bibitem[{Misiriotis {et~al.}(2006)Misiriotis, Xilouris, Papamastorakis,
  Boumis, \& Goudis}]{misiriotis2006}
Misiriotis, A., Xilouris, E.~M., Papamastorakis, J., Boumis, P., \& Goudis,
  C.~D. 2006, A\&A, 999, 999

\bibitem[{{Murphy} {et~al.}(2005){Murphy}, {Armus}, {Helou}, {Braun}, \& {the
  SINGS team}}]{murphy2005}
{Murphy}, E.~J., {Armus}, L., {Helou}, G., {Braun}, R., \& {the SINGS team}.
  2005, ArXiv Astrophysics e-prints

\bibitem[{{Nakai} \& {Kuno}(1995)}]{nakai_kuno1995}
{Nakai}, N. \& {Kuno}, N. 1995, \pasj, 47, 761

\bibitem[{Nakai {et~al.}(1994)Nakai, Kuno, Handa, \& Y.~Sofue}]{nakai1994}
Nakai, N., Kuno, N., Handa, T., \& Y.~Sofue, Y. 1994, PASJ, 46, 527

\bibitem[{Nikola {et~al.}(2001)Nikola, Geis, Herrmann, Madden, Poglitsch,
  Stacey, \& Townes}]{nikola2001}
Nikola, T., Geis, N., Herrmann, F., {et~al.} 2001, ApJ, 561, 203

\bibitem[{Patrikeev {et~al.}(2006)Patrikeev, Fletcher, Stepanov, Beck,
  Berkhuijsen, Frick, \& Horellou}]{patrikeev2006}
Patrikeev, I., Fletcher, A., Stepanov, R., {et~al.} 2006, A\&A, 999, 999

\bibitem[{{Rand} \& {Kulkarni}(1990)}]{rand_kulkarni1990}
{Rand}, R.~J. \& {Kulkarni}, S.~R. 1990, \apjl, 349, L43

\bibitem[{Rots(1980)}]{rots1980}
Rots, A. 1980, A\&A, 41, 189

\bibitem[{Rots {et~al.}(1990)Rots, Crane, Bosma, Athanassoula, \& van~der
  Hulst}]{rots1990}
Rots, A., Crane, P., Bosma, A., Athanassoula, E., \& van~der Hulst, J. 1990,
  Astronomical Journal, 100, 387

\bibitem[{{Rydbeck} {et~al.}(2004){Rydbeck}, {Thomasson}, {Aalto}, {Johansson},
  \& {H{\"u}ttemeister}}]{rydbeck2004}
{Rydbeck}, G., {Thomasson}, M., {Aalto}, S., {Johansson}, L., \&
  {H{\"u}ttemeister}, S. 2004, in ASP Conf. Ser. 320: The Neutral ISM in
  Starburst Galaxies, ed. S.~{Aalto}, S.~{Huttemeister}, \& A.~{Pedlar}, 152

\bibitem[{{Schinnerer} {et~al.}(2004){Schinnerer}, {Wei{\ss}}, {Aalto}, \&
  {Scoville}}]{schinnerer2004}
{Schinnerer}, E., {Wei{\ss}}, A., {Aalto}, S., \& {Scoville}, N.~Z. 2004, in
  The Dense Interstellar Medium in Galaxies, ed. S.~{Pfalzner}, C.~{Kramer},
  C.~{Staubmeier}, \& A.~{Heithausen}, 117

\bibitem[{{Schmidt}(1959)}]{schmidt1959}
{Schmidt}, M. 1959, \apj, 129, 243

\bibitem[{{Schuster} {et~al.}(2004){Schuster}, {Boucher}, {Brunswig}, {Carter},
  {Chenu}, {Foullieux}, {Greve}, {John}, {Lazareff}, {Navarro}, {Perrigouard},
  {Pollet}, {Sievers}, {Thum}, \& {Wiesemeyer}}]{schuster2004}
{Schuster}, K.-F., {Boucher}, C., {Brunswig}, W., {et~al.} 2004, \aap, 423,
  1171

\bibitem[{{Scoville} \& {Young}(1983)}]{scoville_young1983}
{Scoville}, N. \& {Young}, J.~S. 1983, \apj, 265, 148

\bibitem[{{Scoville} {et~al.}(2001){Scoville}, {Polletta}, {Ewald}, {Stolovy},
  {Thompson}, \& {Rieke}}]{scoville2001}
{Scoville}, N.~Z., {Polletta}, M., {Ewald}, S., {et~al.} 2001, \aj, 122, 3017

\bibitem[{{Sofue}(1996)}]{sofue1996}
{Sofue}, Y. 1996, \apj, 458, 120

\bibitem[{{Strong} {et~al.}(1988){Strong}, {Bloemen}, {Dame}, {Grenier},
  {Hermsen}, {Lebrun}, {Nyman}, {Pollock}, \& {Thaddeus}}]{strong1988}
{Strong}, A.~W., {Bloemen}, J.~B.~G.~M., {Dame}, T.~M., {et~al.} 1988, \aap,
  207, 1

\bibitem[{{Strong} \& {Mattox}(1996)}]{strong_mattox1996}
{Strong}, A.~W. \& {Mattox}, J.~R. 1996, \aap, 308, L21

\bibitem[{Takats \& Vinko(2006)}]{takats2006}
Takats, K. \& Vinko, J. 2006, \mnras, 999, 999

\bibitem[{{Thronson} \& {Telesco}(1986)}]{thronson1986}
{Thronson}, H.~A. \& {Telesco}, C.~M. 1986, \apj, 311, 98

\bibitem[{Tilanus \& Allen(1991)}]{tilanus_allen1991}
Tilanus, R. \& Allen, R. 1991, A\&A, 244, 8

\bibitem[{{Tilanus} \& {Allen}(1991)}]{tilanus1991}
{Tilanus}, R.~P.~J. \& {Allen}, R.~J. 1991, \aap, 244, 8

\bibitem[{{Toomre}(1964)}]{toomre1964}
{Toomre}, A. 1964, \apj, 139, 1217

\bibitem[{Toomre \& Toomre(1972)}]{toomre_toomre1972}
Toomre, A. \& Toomre, J. 1972, ApJ, 178, 623

\bibitem[{{Tully}(1974{\natexlab{a}})}]{tully1974b}
{Tully}, R.~B. 1974{\natexlab{a}}, \apjs, 27, 437

\bibitem[{{Tully}(1974{\natexlab{b}})}]{tully1974c}
{Tully}, R.~B. 1974{\natexlab{b}}, \apjs, 27, 449

\bibitem[{{Wong} \& {Blitz}(2002)}]{wong_blitz2002}
{Wong}, T. \& {Blitz}, L. 2002, \apj, 569, 157

\bibitem[{{Yun} {et~al.}(2001){Yun}, {Reddy}, \& {Condon}}]{yun2001}
{Yun}, M.~S., {Reddy}, N.~A., \& {Condon}, J.~J. 2001, \apj, 554, 803

\end{thebibliography}
   
\clearpage 
   
\end{document}